\documentclass[prb,twocolumn,showpacs,showkeys,groupedaddress,amsmath,amssymb]{revtex4}% Physical Review B
\usepackage {amssymb}
\usepackage {amsmath}
\usepackage{graphics,amssymb}
\usepackage{natbib}
\usepackage{revsymb}

\usepackage{graphicx}% Include figure files
\usepackage{dcolumn}% Align table columns on decimal point
\usepackage{bm}% bold math
\newcommand{\DS}{\displaystyle}

\begin{document}

\title{Mechanisms of arsenic clustering in silicon\\}
% Force line breaks with \\

\author{F. F. Komarov}
\affiliation{Department of Physical Electronics, Belarusian State
University, 1 Kurchatov
Str., Minsk 220064, Belarus \\}%
%e-mail: KomarovF@bsu.by

\author{O. I. Velichko} %\email[]{oleg_velichko@lycos.com}
\affiliation{Department of Physics, Belarusian State University on
Informatics and Radioelectronics, 6 P. Brovka Str., Minsk 220013,
Belarus
\\}

\author{V. A. Dobrushkin} \email[]{dobrush@dam.brown.edu}
\affiliation{Brown University, Division of Applied Mathematics,
Providence, Rhode Island 02912, USA
\\}

\author{A. M. Mironov}
\affiliation{Institute of Applied Physics Problems, 7 Kurchatov
Str., Minsk 220064, Belarus
\\}
% e-mail: MironovA@bsu.by

\begin{abstract}

A model of arsenic clustering in silicon is proposed and analyzed.
The main feature of the proposed model is the assumption that
negatively charged arsenic complexes play a dominant role in the
clustering process. To confirm this assumption, electron density
and concentration of impurity atoms incorporated into the clusters
are calculated as functions of the total arsenic concentration at
a temperature of 1050\,$^{\circ}$C. A number of the negatively
charged clusters incorporating a point defect and one or more
arsenic atoms $(\text{DAs}_{1})^{-}$, $(\text{DAs}_{1})^{2-}$,
$(\text{DAs}_{2})^{-}$, $(\text{DAs}_{2})^{2-}$,
$(\text{DAs}_{3})^{-}$, $(\text{DAs}_{3})^{2-}$,
$(\text{DAs}_{4})^{-}$, and $(\text{DAs}_{4})^{2-}$ are
investigated. It is shown that for the doubly negatively charged
clusters or for clusters incorporating more than one arsenic atom the
electron density reaches a maximum value and then monotonically
and slowly decreases as total arsenic
concentration increases. In the case of cluster $(\text{DAs}_{2})^{2-}$, the
calculated electron density agrees well with the experimental
data. Agreement with the experiment confirms the conclusion that
two arsenic atoms participate in the cluster formation. Among all
present models, the proposed model of clustering by formation of
$(\text{DAs}_{2})^{2-}$ gives the best fit to the experimental
data and can be used in simulation of high concentration arsenic
diffusion.
\end{abstract}

% PACS
%Doping and impurity implantation in germanium and silicon                             % Classification Scheme.


%Use showkeys class option if keyword
                              %display desired

\maketitle

\section{Introduction}

Using low energy high dose arsenic ion implantation, one can
produce the active regions of modern integrated microcircuits
characterized by very shallow junctions and high dopant
concentrations. Thermal annealing is applied after implantation
for the arsenic activation and damage reduction. During the
initial stage of annealing, the arsenic atoms become electrically
active, occupying the substitutional positions. Then a fraction of
impurity atoms incorporates into the clusters, thereby decreasing
the layer conductivity.
\cite{Tsukamoto_80,Tsai_80,Fahey_89,Nobili_94,
Solmi_98,Krishnamoorthy_98,Nobili_99,Uematsu_00,
Nobili_01,Solmi_01,Solmi_03} Due to clustering, the concentration
of charge carriers is less than the total impurity concentration
because the clustered dopant atoms do not serve as a source of
free charge carriers. We can observe the clustering phenomenon in
many cases of the formation of highly doped semiconductor
structures; for example, in the region with high impurity
concentration during thermal diffusion of arsenic
atoms.\cite{Fair_73,Fair_Weber_73,Murota_79} Deactivation of the
electrically active dopant atoms also indicates that clustering
can occur during thermal treatment of the supersaturated arsenic
layers created by ion implantation with subsequent thermal or
laser annealing.
\cite{Nobili_94,Nobili_99,Parisini_90,Luning_92,Rousseau_94,
Rousseau_98,Solmi_02} Moreover, a reverse annealing or transient
increase of the carrier density can be observed if a doped
specimen, previously annealed at temperature $T_{1}$, is further
annealed at higher temperature $T_{2}$. \cite{Solmi_00} It is
supposed that transient dissolution of a part of the clusters
occurs during such a reverse annealing. \cite{Solmi_00}

At very high dopant concentrations exceeding the solid solubility
of As in Si, a significant fraction of arsenic atoms can form
precipitates.
\cite{Fahey_89,Nobili_94,Solmi_98,Krishnamoorthy_98,Uematsu_00,Solmi_01,
Parisini_90, Nobili_83,La Via_91} The typical distributions of the
total dopant concentration and electron density after thermal
annealing of silicon substrates heavily implanted by As were
presented by Nobili et al. \cite{Nobili_94} (Fig. 7) and by Solmi
\cite{Solmi_01} (Fig. 1c). As can be seen from Fig. 1c,
\cite{Solmi_01} the carrier concentration reaches its saturation
value $n_{e}=3.57\times10^{8}$ $\mu $m$^{{\rm - }{\rm 3}}$,
whereas the total arsenic concentration increases to the
solubility limit $C_{sol}=3.27\times10^{9}$ $\mu $m$^{{\rm - }{\rm
3}}$ \cite{Nobili_01} and continues to increase in the region
adjoint to the surface. It is supposed that the cluster formation
occurs at high dopant concentrations,
\cite{Tsukamoto_80,Tsai_80,Fahey_89,Nobili_99,Nobili_01,Solmi_01,Fair_73,
Parisini_90,Luning_92} more precisely, at the dopant
concentrations approximately ranging from $n_{e}$ to $10\,n_{e}$.
If dopant concentration exceeds the solubility limit $C_{sol}$,
precipitation occurs. \cite{Nobili_94,Solmi_01} Arsenic clustering
and precipitation have attracted evergrowing attention of the
researchers, as they are greatly important for the silicon
integrated circuits technology. Both clusters and precipitates do
not serve as a source of charge carriers and the total arsenic
concentration may be higher than the electron density by about two
orders of magnitude. The clusters and precipitates are metastable
and can dissociate under subsequent thermal treatments. This means
that stringent requirements should be placed upon the accuracy of
the clustering models used to calculate the carrier concentrations
in heavily doped silicon layers.

\section{Analysis of clustering phenomena}

The development of a model for As clustering necessitates a
consistent analysis of the experimental data and theoretical
substantiation of the processes proceeding in heavily doped
silicon layers.

\subsection{Clustering during thermal diffusion
and annealing of  ion-implanted layers}

A great number of clustering models with different types of
clusters have been developed and proposed since 1973.
\cite{Tsukamoto_80,Tsai_80,Fahey_89,Fair_73,Parisini_90,Luning_92,
Guerrero_82,Berding_98,Mueller_03} The variety of the proposed
models reflects discrepancies in the experimental results
\cite{Nobili_83} and a great uncertainty in the nature of the
clusters discussed. For example, inactive cluster $\text{VAs}_{2}$
was considered in \cite{Fair_73} to explain the difference between
the total and electrically active dopant concentrations for
arsenic thermally diffused from a constant source. For the best
fit to the diffusion profiles of ion-implanted arsenic, it was
suggested that the electrically inactive arsenic atoms were
incorporated into the clusters with four arsenic atoms per
cluster. \cite{Tsukamoto_80} According to Tsai et al.
\cite{Tsai_80} who have also investigated diffusion in the
ion-implanted layers, clustering of arsenic may be expressed as

\begin{equation} \label{Cluster_Reaction_Tsai}
3\text{As}^{+}+e^{-}{\stackrel{\mbox{Annealing}}{\qquad
\longleftrightarrow \qquad }}
\text{As}_{3}^{2+}{\stackrel{\mbox{25\,$^{\circ}$C}}{\quad
\longrightarrow \quad }} \text{As}_{3},
\end{equation}

\noindent where $\text{As}^{ +} $  is the substitutionally
dissolved arsenic atom participating in the cluster formation;
$e^{ - }$ is the electron.

As follows from reaction (\ref{Cluster_Reaction_Tsai}), the
clustered As atoms are electrically active at annealing
temperatures, but are neutral at room temperature.

It was found in \cite{Uematsu_00} that the formation of clusters
$(\text{As}_{2}\text{V})^{\times}$ occurs due to the following
reaction, which

\begin{equation} \label{Cluster_Reaction_Uematsu}
2\text{As}^{+}+\text{V}^{2-}\longleftrightarrow
(\text{As}_{2}\text{V})^{\times}
\end{equation}

\noindent is most liable to cause deactivation because the
concentration of inactive As species increases to the fourth power
of the active As concentration. \cite{Tsukamoto_80,Murota_79} Here
$\text{V}^{2-}$ is the doubly negatively charged vacancy
participating in the clustering.

Indeed, if a local equilibrium is assumed, it follows from the mass action
law for reaction (\ref{Cluster_Reaction_Uematsu}) that

\begin{equation} \label{MAL_Uematsu}
C^{A}=A\tilde{C}^{V\times}\chi^{2}C^{2},
\end{equation}

\begin{equation} \label{Constant_Uematsu}
A=H^{A} C_{i}^{V\times} n_{i}^{2},
\end{equation}

\begin{equation} \label{Rel_vacancy_concentration}
\tilde{C}^{V\times}=\frac{C^{V\times}}{C_{i}^{V\times}},
\end{equation}

\noindent where $C^{A}$ and $C$ are the concentrations of the
impurity atoms incorporated into clusters and substitutionally
dissolved arsenic atoms, respectively; $H^{A}$ is the constant of
the local equilibrium for reaction
(\ref{Cluster_Reaction_Uematsu}); $C^{V\times}$ and
$C_{i}^{V\times}$ are the actual and the equilibrium
concentrations of neutral vacancies, respectively;
$\chi=\displaystyle \frac{n}{n_{i}}$ is the electron density
normalized to the concentration of intrinsic charge carriers in a
semiconductor during thermal treatment $n_{i}$.

One would expect from (\ref{MAL_Uematsu}) that reaction
(\ref{Cluster_Reaction_Uematsu}) yields the fourth power
dependence $C^{A}\propto C^{4}$, since for high concentrations
$C>>n_{i}$ it might be reasonable to assume $\displaystyle
\chi=\frac{n}{n_{i}}\thickapprox \frac{C}{n_{i}}$. However, as
seen from Fig. 7 (Ref. \cite{Nobili_94}) and Fig. 1c (Ref.
\cite{Solmi_01}), the electron density $n\thickapprox n_{e}=const$
in the region of concentrations associated with active clustering.
Consequently, reaction (\ref{Cluster_Reaction_Uematsu}) yields the
second power dependence $C^{A}\backsim C^{2}$, which  contradicts
the experimental data. \cite{Tsukamoto_80,Murota_79}

Because of the vacancy participating in cluster formation, silicon
self-interstitials were in turn ejected or left behind during
clustering. \cite{Krishnamoorthy_98} This assumption had been made
much earlier in the study \cite{Parisini_90,Rousseau_94} of
impurity deactivation within the laser-annealed layers. However,
the results obtained in \cite{Krishnamoorthy_98} show that the
point defects induced during clustering and/or precipitation make
no contribution to the enhanced transient diffusion of arsenic
implanted at low energy, while these defects did coalesce to form
extended defects at the projected range. Analysis of the defect
microstructure has revealed that the transition between arsenic
clustering and SiAs precipitation is not abrupt, pointing to
possible coexistence of arsenic clusters and SiAs precipitates
(although precipitates are not directly observed).
\cite{Krishnamoorthy_98}

The model \cite{Tsai_80} was generalized by Solmi and Nobili
\cite{Solmi_98}

\begin{equation} \label{Cluster_Reaction_Solmi}
m\text{As}^{+}+e^{-}{\stackrel{\mbox{Annealing}}{\qquad
\longleftrightarrow \qquad }}
\text{As}_{m}^{(m-1)+}{\stackrel{\mbox{25\,$^{\circ}$C}}{\quad
\longrightarrow \quad }} \text{As}_{m}
\end{equation}
\noindent to take into account a saturation behavior of the
carrier density with increasing dopant concentration. For
diffusion at a temperature of 1050\,$^{\circ}$C, the best agreement
between the modeled and experimental curves describing diffusivity
against total arsenic concentration was achieved for $m=4$. And at
900\,$^{\circ}$C, the neutral cluster model \cite{Fair_73} provided
a better agreement with the experimental data. The reaction
(\ref{Cluster_Reaction_Solmi}) was also used in \cite{Nobili_01}
to describe deactivation in the arsenic-doped layers, which were
formed by ion implantation and high temperature annealing of
silicon-on-insulator specimens. However, Solmi et al.
\cite{Solmi_03} have recently used the following reaction

\begin{equation} \label{Cluster_Reaction_Solmi_03}
m\text{As}^{+}+\text{V}\longleftrightarrow \text{As}_{m}+\text{I}
\end{equation}

\noindent to model the transient arsenic diffusion. Here
$m$ is assumed to have the values between 2 and 4 to take into
account the fact that the As cluster was formed around a vacancy with
the subsequent injection of the self-interstitial.

It is important to note that reaction
(\ref{Cluster_Reaction_Solmi}) was first theoretically studied by
Guerrero et al. in paper, \cite{Guerrero_82} which will be
considered below.

\subsection{Clustering during deactivation of supersaturated
arsenic layers formed by laser annealing}

Let us consider the clustering models used for the explanation of
deactivation in the supersaturated arsenic layers formed by ion
implantation and subsequent laser annealing. In \cite{Luning_92}
it was found that a cluster consisting of four positively charged
arsenic atoms and two additional negative charges gave the best
least-squares fit with respect to the experimental data obtained
in the case of deactivation within the supersaturated arsenic
layers created by ion implantation and subsequent laser annealing.
If two negative charges are associated with a doubly charged
vacancy, this model gains support from the theoretical
calculations of Pandey et al. \cite{Pandey_88} Such a cluster has
an intuitive geometrical and structural appeal --- four arsenic
atoms tetrahedrally arranged around a vacancy may relax inward,
thus relieving lattice strains due to the As size effect.
\cite{Luning_92} Using this cluster species, a single-step
clustering reaction was postulated in \cite{Luning_92}

\begin{equation} \label{Cluster_Reaction_Luning}
4\text{As}^{+}+\text{V}^{2-}\longleftrightarrow
(\text{As}_{4}\text{V})^{2+}.
\end{equation}

The assumption of self-interstitial ejection during clustering was
supported by the experimental data obtained by Parisini et al.
\cite{Parisini_90} who have observed that extended defects at the
projected range were extrinsic or interstitial in nature. The reaction

\begin{equation} \label{Cluster_Reaction_Parisini}
(\text{As}_{2})^{2+}+2e^{-}\longrightarrow
(\text{As}_{2}\text{V})^{+}+\text{I}^{-}
\end{equation}

\noindent was proposed in \cite{Parisini_90} only for the initial
step of the deactivation process in arsenic implanted silicon
specimens, first laser-annealed and then thermally annealed. Here
$\text{I}^{-}$ is the negatively charged self-interstitial. The
results obtained by Rousseau et al. \cite{Rousseau_94} confirm the
conclusion that arsenic is indeed deactivated by vacancies, with a
concurrent injection of self-interstitials. The latter follows
from the observed significant enhancement of the diffusion in the
buried boron layer underneath the As structure subjected to
deactivation. Investigations of the enhanced diffusion in buried
boron layers were continued in, \cite{Rousseau_98} where it was
proposed that small arsenic clusters of various sizes were formed
around a vacancy during deactivation with injection of the
associated interstitial into the bulk.

Considering that the majority of the models described in this
section are based essentially on the experimental investigations
of the defect subsystem, we first concern ourselves with these
experimental results.

\subsection{Experimental investigations of defect subsystem}

The methods of transmission electron spectroscopy (TEM),
\cite{Nobili_94,Krishnamoorthy_98,Parisini_90,La
Via_91,Dokumachi_95} extended x-ray-absorption fine-structure
(EXAFS), \cite{Parisini_90,Erbil_86,Allain_92,d'Acapito_04}
Rutherford backscattering (RBS), \cite{La Via_91,Brizard_93}
positron lifetime measurements
\cite{Makinen_89,Lawther_95,Myler_96,Szpala_96} combined with the
measurements of electron momentum distributions,
\cite{Polity_98,Saarinen_99,Ranki_03,Ranki_Saarinen_03,Ranki_04}
and other methods \cite{Nobili_83,Herrera-Gomez_96} are commonly
used to investigate silicon specimens doped by As. For example,
TEM observations show that As-lased samples are completely free of
any visible defects. \cite{Dokumachi_95} On the other hand, TEM
analysis reveals the presence of extended defects after subsequent
thermal treatment. \cite{Parisini_90,Dokumachi_95} For example,
small \{113\} interstitial loops begin to appear beyond the
plateau region of the As profile after low-temperature treatment.
\cite{Parisini_90} After high-temperature thermal treatment
($>$850\,$^{\circ}$C), a low density of very small As-related
precipitates (about 2 nm in diameter) is observed. The
precipitated As fraction cannot be responsible for the total
amount of inactive As. \cite{Parisini_90}

It was established that EXAFS measurements provided particular
details about the local atomic structure around the dopant and
confirmed the fact that As atoms are substitutional after laser
annealing. \cite{Erbil_86,Allain_92} Note that, according to the
x-ray standing-wave spectroscopy measurements, the As atoms
remained at the substitutional positions even after 85\% of the
electrical activity has been lost due to thermal annealing of the
laser melted layers. \cite{Herrera-Gomez_96} On the other hand, a
large amount of nonsubstitutional arsenic was detected by
Rutherford back-scattering after diffusion from $\text{TiSi}_{2}$
source. \cite{La Via_91} These defects are interpreted as
precipitates, probably formed due to the stress induced by
$\text{TiSi}_{2}$ layer. Indeed, when thermal annealing is carried
out after the complete removal of titanium silicide (i.e., without
stress), one can observe the activation of all the arsenic atoms
diffused in the silicon substrate. \cite{La Via_91} Thus, such
full activation indicates the significant influence of stress on
the formation of precipitates.
%
% very important for future analysis
%

According to EXAFS measurements, \cite{Allain_92} subsequent
thermal annealing (350\,\,--\,750\,$^\circ $C) of the As-lased
layers leads to the formation of $VAs_{m}$ clusters including up
to $7\pm4$ As atoms around the vacancy.  However, at high
temperatures it was observed that the number of the first
neighbors of Si has tended back to 4 atoms.
%
% very important for this analysis
%
This is probably due to the precipitates observed by TEM over
this temperature range. Quantitative analysis of EXAFS data for
the ultra-low energy implanted layers after rapid or spike
annealing \cite{d'Acapito_04} has revealed a site for As that was
different from the pure substitutional one, suggesting the presence
of clusters of As atoms coupled to vacancies. All the observed
phenomena, namely, a low value of the total coordination number in
the annealed samples, small $\text{As--Si}$ bond length, and the
presence of $\text{As--As}$ coordinations, may  be explained by
the assumption that clustering of As ions with vacancies takes
place to form $\text{VAs}_{m}$ structures.

Ion channeling and RBS were used in \cite{Brizard_93} to
complement the EXAFS and electron microscopy results. In has been
found that the deactivated As atom is displaced with respect to
the lattice sites and, moreover, the average displacement is
constant over the temperature range 450\,\,--\,900\,$^{\circ}$C,
being equal to $0.23 \pm 0.06$ \AA. Also, in \cite{Brizard_93} it
was proposed  that the formation of larger clusters
$\text{V}_{n}\text{As}_{m}$ ($n\thickapprox 5$ and $m\thickapprox
10$) occurred starting from $\text{VAs}_{2}$ or $\text{VAs}_{4}$.
The formation of these larger clusters is in agreement with a
decrease in the first nearest neighbors of As. At higher
temperatures, an increase in the number of the first nearest
neighbors of As to 4 was observed by EXAFS measurements.
\cite{Allain_92} This means that the cluster species may differ at
intermediate (350\,\,--\,750\,$^{\circ}$C) and higher
temperatures.

Very interesting information about clustering was obtained during
the experimental investigations of the doped layers by  positron
annihilation spectroscopy
\cite{Lawther_95,Myler_96,Saarinen_99,Ranki_03,Ranki_Saarinen_03,Ranki_04}
and by positron annihilation spectroscopy combined with the
Hall-effect/resistivity measurements. \cite{Ranki_03} The defect
subsystem was generally investigated before and after thermal
treatment. To support these measurements, the positron lifetimes
and core electron momentum distributions were calculated for
different vacancy-donor complexes. \cite{Ranki_04} Lawther et al.
\cite{Lawther_95} have investigated the doped layers melted by an
excimer laser to obtain the profile with constant As concentration
and a sharp fall-off at a depth of 0.2 $\mu$m. Arsenic
deactivation was initiated by annealing at 750\,$^{\circ}$C for 15
s or by conventional thermal treatment for 2 h. It was found that
$\text{VAs}_{m}$ complexes with the average values of $m$ greater
than 2 caused arsenic  deactivation in heavily doped Si. Myler et
al. \cite{Myler_96} have studied the silicon layers (with arsenic
concentration 4$\times 10^{8}$ $\mu \text{m}^{3}$) fully activated
by laser melting and subsequently annealed for 15 s at 500 and
750\,$^{\circ}$C. The changes in the positron annihilation spectra
after the thermal treatment at 500 or 750\,$^{\circ}$C were also
attributed to the formation of
$\text{As}_{m}\text{Si}_{4-m}\text{V}_{ac}$ complexes. In this
experiment, the impossibility of determining the number of
impurity atoms incorporated in the impurity-vacancy complex was
established. Ranki at al. \cite{Ranki_03} have studied the samples
implanted by As during MBE growth at 450\,$^{\circ}$C. The samples
were annealed in $\text{N}_{2}$, either by RTA at 900\,$^{\circ}$C
for 10\,\,--\,170 s or furnace annealing at
800\,\,--\,900\,$^{\circ}$C for two minutes. Based on the positron
annihilation and Hall-effect/resistivity experiments, Ranki et al.
\cite{Ranki_03} have concluded that the dominant defect in
As-grown samples was $\text{VAs}_{3}$. The measurements
demonstrated high concentrations (above $10^{5}$ $\mu
\text{m}^{-3}$) of $\text{VAs}_{3}$ complexes in heavily-doped
silicon. Moreover, larger V-As complexes, probably
$\text{V}_{2}\text{As}_{5}$, may occur together with
$\text{VAs}_{3}$ at high As concentrations. A relative amount of
$\text{V}_{2}\text{As}_{5}$ increases with annealing. This cluster
is even dominant after annealing at 800\,$^{\circ}$C. The
$\text{VAs}_{3}$ and $\text{V}_{2}\text{As}_{5}$ complexes become
unstable at 800\,$^{\circ}$C and 900\,$^{\circ}$C, respectively,
then their concentrations decrease. Cluster reconstruction may
occur during cooling , starting from $\text{VAs}_{1}$ to
$\text{VAs}_{2}$ and then to $\text{VAs}_{3}$. The clusters
$\text{VAs}_{2}$ and $\text{VAs}_{3}$ may be in turn transformed
to the $\text{V}_{2}\text{As}_{5}$ complexes. Moreover, some
$\text{VAs}_{4}$ and other vacancy complexes may be also formed,
but with much lower concentrations.

High energy electron irradiation is commonly used in positron
experiments for the generation of nonequilibrium vacancies, mobile
even at room temperature and liable to interact with the dopant
atoms. \cite{Saarinen_99,Ranki_02,Ranki_Saarinen_03,Ranki_04} For
example, positron experiments \cite{Saarinen_99} were performed
both with the As-grown samples doped by arsenic in concentrations
of $10^{7}$ and $10^{8}$ $\mu \text{m}^{-3}$ and samples subjected
to 2 MeV electron irradiation at 300 K. It was established that
heavily As-doped silicon contained $\text{VAs}_{3}$ complex as a
native defect. Before irradiation, the concentration of
$\text{VAs}_{3}$ was equal to $\sim 10^{7}$ $\mu \text{m}^{-3}$ at
a doping level of $10^{8}$ $\mu \text{m}^{-3}$. After electron
irradiation, one can observe the pairs ($\text{VAs}_{1}$) formed
by a vacancy and a single impurity atom.  It was demonstrated in
\cite{Ranki_02} that the migration of these $\text{VAs}_{1}$
defects started at around 450 K, leading to formation of
$\text{VAs}_{2}$ defects. And, in turn, these defects were
transformed to $\text{VAs}_{3}$ defects at 700 K. The
$\text{VAs}_{3}$ defects were stable at 700 K, representing the
dominant vacancy-impurity cluster in heavily doped n-type Si at
this temperature. The formation of larger $\text{VAs}_{2}$ and
$\text{VAs}_{3}$ complexes was significantly dependent on the As
concentration. This interpretation of the experimental data was
confirmed by the results obtained in. \cite{Ranki_Saarinen_03} In
these experiments, the electron irradiated samples were annealed
isochronally (30 min) at 300\,\,--\,1220 K. Irradiation of the
heavily doped Si samples has produced mainly the vacancy-donor
pairs ($\text{VAs}_{1}$) with a small concentration of
divacancies. Considering that, after irradiation, $\text{VAs}_{1}$
concentration was much higher than the initial concentration of
$\text{VAs}_{3}$, no signals from $\text{VAs}_{3}$ were observed
for the As-irradiated As-doped sample. From the core-region
electron momentum distribution measurements, it was found
\cite{Ranki_Saarinen_03} that defects in As-irradiated samples may
be identified as $\text{VAs}_{1}$, in samples annealed at 600 K
--- as $\text{VAs}_{2}$, and in those annealed at 775 K --- as
$\text{VAs}_{3}$. It is well known that a drastic drop in the
conductivity occurs when heavily doped Si is annealed at
temperatures between 400\,$^{\circ}$C and 500\,$^{\circ}$C. This
deactivation of the dopants is partially reversible by annealing
at 800\,\,--\,1000\,$^{\circ}$C. Taking into account these
experimental data, Ranki et al. \cite{Ranki_Saarinen_03} have
suggested that the formation and annealing of $\text{VAs}_{3}$ at
700 and 1100 K, respectively, are responsible for the observed
behavior of the conductivity. The experiments conducted in
\cite{Ranki_02} were completed by studies in \cite{Ranki_04}
associated with thermal treatment up to 1220 K. It has been found
that dissociation of $\text{VAs}_{3}$ began at 1100 K, and
at 1220 K these defects were annealed away.

Thus, the analyzed data show that complexing of As atoms with
vacancies occurs in the layers heavily doped by arsenic. Based on
the experimental data, various $\text{V}_{m}\text{As}_{n}$
clusters are possible but $\text{VAs}_{2}$ and $\text{VAs}_{3}$
are the most likely. Taking clustering into account, one can
explain the phenomenon of compensation at high doping levels.
Nevertheless, some aspects are not clearly understood. For
example, $\text{VAs}_{3}$ starts to dissociate at 1100 K. This
annealing temperature agrees well with the observation that the
number of Si first neighbors tends back to 4 atoms
\cite{Allain_92} at temperatures higher than 750\,$^{\circ}$C. At
the same time, clustering occurs at temperatures higher than
750\,$^{\circ}$C as well. \cite{Solmi_01} This means that such a
high temperature clustering is hardly explained by
$\text{VAs}_{3}$ complexes.

% grown by MBE and doped layers irradiated by electrons

\subsection{First principles studies}

Along with various experimental investigations, a number of the
theoretical {\it ab initio} calculations have been performed to
explain deactivation of arsenic atoms in silicon. From the
calculations of Pandey et al., \cite{Pandey_88} it follows that
$\text{VAs}_{4}$ complex including a vacancy surrounded by four
arsenic atoms is energetically favored over both substitutional,
isolated As in Si and substitutional $\text{Si\,-As}_{4}$
configurations. This cluster is electrically inactive, being
responsible for arsenic deactivation and structural changes in
heavily doped silicon. Larger defect clusters (e.g.,
$\text{V}_{2}\text{As}_{4}$) should also form during annealing,
whereas $\text{VAs}_{4}$ is only the first step in the clustering
process. These investigations were continued in,
\cite{Ramamoorthy_96} where general $\text{V}_{n}\text{As}_{m}$
complexes have been considered. The formation energies of the
vacancy-donor complexes $\text{VAs}_{1}$, $\text{VAs}_{2}$,
$\text{VAs}_{3}$, and $\text{VAs}_{4}$ were found to be 2.47,
0.82, -0.53, and -2.39 eV, respectively. Moreover, it was proposed
that the complex $\text{VAs}_{2}$ was mobile, as were the
$\text{VAs}_{1}$ pairs. As these pairs moved, they reacted with
other defects to form larger, immobile complexes. When considering
the mobile $\text{VAs}_{2}$ complexes, Ramamoorthy and Pantelides
\cite{Ramamoorthy_96} have tried to explain the coupled diffusion
phenomenona and clustering. It was supposed that $\text{VAs}_{2}$
and $\text{VAs}_{3}$ were the dominant complexes in the
deactivated specimens near the enhanced-diffusion threshold.
\cite{Larsen_86,Larsen_93} A high rate of As diffusion observed
during rapid thermal annealing \cite{Larsen_86} was due to
$\text{VAs}_{2}$ migration over a very short time period
(6\,\,-\,60 s), when no extensive clustering could occur. During
their motion, $\text{VAs}_{2}$ complexes reacted with other
defects and formed larger, immobile complexes, which decreased the
As diffusivity.

Nevertheless,  further investigation is necessary because some
theoretical results disagree with the experimental data. For
example, in \cite{Chadi_97,Citrin_03} it was stated that the
formation of $\text{VAs}_{3}$ or $\text{VAs}_{4}$ defects was
exothermic, but, according to the first-principles calculations,
\cite{Ramamoorthy_96} buildup of $\text{VAs}_{1}$ or
$\text{VAs}_{2}$ structures was endothermic. This means that even
at low arsenic concentrations all impurity atoms must form
$\text{VAs}_{3}$ or $\text{VAs}_{4}$ clusters after long-time
treatment. In point of fact, at low dopant concentrations, one can
observe intensive diffusion by means of $\text{VAs}_{1}$ pairs
rather than clustering.

Berding et al., \cite{Berding_98} and Berding and Sher
\cite{Berding_Sher_98} have used the electronic quasichemical
formalism to calculate a free energy of various clusters,
including $\text{As\,Si}_{4}$, $\text{VAs}_{4}$,
$\text{VAs}_{3}\text{Si}_{1}$, $\text{VAs}_{2}\text{Si}_{2}$. Here
$\text{As\,Si}_{4}$ is the arsenic atom in the substitutional
position. The neutral cluster composed of threefold-coordinated
second-neighbor arsenic atoms DP(2) was proposed by Chadi et al.
\cite{Chadi_97} as an alternative to exothermic $\text{VAs}_{4}$.
The clusters DP(2), $\text{V}_{2}\text{As}_{6}$, and
$\text{Si}_{8}$ were also included in the calculations performed
in. \cite{Berding_Sher_98} In contrast to, \cite{Ramamoorthy_96}
Berding et al., \cite{Berding_98} Berding and Sher
\cite{Berding_Sher_98} take into account the ionized states and
entropy of the cluster formation. The entropy is unfavourable for
the formation of a large defect complex such as $\text{VAs}_{4}$.
Therefore, the complete free-energy calculation is needed to
determine the role of $\text{VAs}_{4}$ in deactivation. Based on
the complete free-energy calculation, it was found that cluster
$\text{VAs}_{4}$ was neutral, in agreement with the previous
findings. \cite{Pandey_88} Also, the energy of this complex (-1.62
eV) was in a rough agreement with the value given in.
\cite{Pandey_88} The $\text{VAs}_{3}$ and $\text{VAs}_{2}$
clusters were found to have one and two acceptor levels within the
energy gap. Consequently, their formation energy may be
effectively decreased when the Fermi energy is near the
conduction-band edge. The equilibrium cluster concentrations
depending on the temperature and concentration of dopant atoms are
obtained using the minimized free energy of the system. For all
the dopant concentrations and temperatures considered
(400\,\,--\,1000\,$^{\circ}$C), three classes of clusters are
dominant under equilibrium conditions: $\text{Si}_{8}$,
$\text{AsSi}_{4}$, and $\text{VAs}_{4}$. In all cases, the Fermi
energy was near the conduction-band edge. Predominantly,
$\text{VAs}_{3}$ and $\text{VAs}_{2}$ were singly and doubly
ionized, respectively. At low arsenic concentrations (up to $\sim
5\times10^{6}$ $\mu$\text{m}$^{-3}$), noticeable deactivation was
absent for temperatures higher than 500\,$^{\circ}$C. With the
arsenic concentration raised to $5\times10^{7}$ $\mu
$\text{m}$^{-3}$ and higher, significant deactivation was
predicted, mainly due to the formation of $\text{VAs}_{4}$
clusters. Both concentration and temperature influenced the
contributions of different defects. For example, the concentration
of $\text{VAs}_{4}$ clusters has reached the concentration of
isolated arsenic atoms in the lattice $\text{As\,Si}_{4}$ for
$\thicksim 5\times10^{8}$ $\mu $\text{m}$^{-3}$ at a temperature
of 700\,$^{\circ}$C and for $\thicksim 2\times10^{9}$ $\mu
$\text{m}$^{-3}$ at a temperature of 1000\,$^{\circ}$C.

In \cite{Berding_98,Berding_Sher_98} the authors have attempted to
explain the effect of the electron concentration saturation at
high doping levels. However, the calculations performed in these
papers show that saturation is not reachable even at very high
arsenic concentrations. For example, saturation was not reached
with the arsenic concentration of $2\times10^{9}$ $\mu $m$^{-3}$
at a temperature of 700\,$^{\circ}$C, which is in disagreement
with the experimental data (see Fig. 2 from \cite{Berding_98}). It
is important that, according to, \cite{Berding_Sher_98} DP(2)
clusters proposed by Chadi et al. in \cite{Chadi_97} may be
present as a deactivating species, insignificantly contributing to
the deactivation under conditions of full equilibration.

These new defects, first mentioned in \cite{Chadi_97} and called
DP, represent a pair of two three-fold coordinated donor atoms.
The lowest energy DP structures are DP(2) and DP(4), where the
donor atoms occupy either second- or fourth-neighbor Si sites
along the $<110>$ direction. It was shown that DP(2) exist in the
stable electrically active 2+ charge state that is donating
electrons to the conduction band or in the metastable neutral
charge state capturing two electrons from the Fermi sea. A very
important conclusion made in \cite{Chadi_97} is a threefold
coordination of each dopant atom in DP(2) in the neutral trap
state. Thus, relaxation of the neighboring Si atoms creates an
increased volume, which is consistent with the experimental
observations. Therefore, such an increased volume around the donor
atoms is possible without vacancies. The main ideas of
\cite{Chadi_97} were further developed in. \cite{Citrin_03} Citrin
et al. \cite{Citrin_03} have proposed a new class of defects,
called the donor-pair-vacancy-interstitials and composed of two
dopant donor atoms near the displaced Si atom that is forming a
vacancy-interstitial pair. This defect, which is like a hybrid of
the donor-pair and Frenkel-pair defects, is denoted DP(i)V-I. For
the case of Sb, the data of annular dark-field (``Z-contrast'')
scanning transmission electron microscopy (ADF-STEM) clearly
demonstrate that, in heavily Sb-doped silicon grown at low
temperatures, the primary deactivating defect contains only two Sb
atoms. \cite{Voyles_02} Thus, energetically favorable
$\text{VSb}_{3}$ and $\text{VSb}_{4}$ may be ruled out from the
deactivation process. Only $\text{VSb}_{2}$, DP(2), and DP(4) may
be considered as liable candidates. Using additional STEM
measurements, x-ray absorption data, and first-principles
calculations, Voyles et al. \cite{Voyles_02} have shown that
neither $\text{VSb}_{2}$ nor DP(2) and DP(4) defects are important
in heavily Sb-doped Si. At that time, DP(2)V-I and DP(4)V-I were
in line with the available experimental results, including
positron annihilation spectroscopy data. \cite{Citrin_03} Both
DP(2)V-I and DP(4)V-I are independent out of the pre-existing
vacancy population, which is in conformity with the observations
of. \cite{Solmi_02}

Nevertheless, as follows from the experimental data,
\cite{Solmi_01} a deactivation mechanism can be different in the
As- and Sb-doped silicon. The mechanism of arsenic clustering
calls for further investigation. In, \cite{Mueller_03} Mueller et
al. have investigated the electronic structure and charge states
of various vacancy-impurity clusters using the first-principles
density-functional theory (DFT). It was found that the
$\text{VAs}_{1}$ complex can trap up to two conduction electrons
at a high n-doping level. Both $\text{VAs}_{2}$ and
$\text{VAs}_{3}$ act as a single-electron trap center. In
agreement with,
\cite{Pandey_88,Ramamoorthy_96,Berding_98,Berding_Sher_98} Mueller
et al. \cite{Mueller_03} have found that in $\text{VAs}_{4}$ the
fifth valence electrons of all four As atoms are strongly bound
and the complex is expected to remain neutral for any position of
the Fermi level. Note that according to the calculations of,
\cite{Berding_98} the cluster $\text{VAs}_{2}$ has two acceptor
levels, whereas Solmi et al. \cite{Solmi_00} have found from the
carrier mobility measurements that the complex to be electrically
neutral at room temperature. In \cite{Xie_99} the first-principles
calculations were performed to explain the coupled diffusion
phenomenona and clustering in heavily arsenic-doped silicon. As
contrast to, \cite{Ramamoorthy_96} it was found that the
$\text{VAs}_{2}$ cluster is less mobile due to its high migration
barrier. However, these clusters can contribute to the diffusion
of As at elevated temperatures.

Proceeding from the presented analysis, our understanding of the
clustering mechanisms is neither complete nor firmly established.
This concerns the number of arsenic atoms in the cluster as well as
its structure. However, a study based on the first-principles
shows that clusters of a certain kind have an acceptor level and can
be negatively charged. Let us consider the models for the clusters
with different charge states.

\subsection{Models based on the mass action law}

The models based on thermodynamic formalism are usually used for
simulation of high concentration dopant diffusion taking place
during semiconductor processing. Assuming a local thermodynamic
equilibrium among substitutionally dissolved arsenic, the dopant
atoms incorporated into clusters, and electrons, one can use the
mass action law to calculate the concentration of clustered dopant
atoms. \cite{Tsai_80,Fair_73,Fair_Weber_73,Guerrero_82} The charge
conservation law for the reaction of cluster formation under the
assumption of local charge neutrality is also useful for
describing the clustering phenomenon. A widespread model of Tsai
et al. \cite{Tsai_80} was the first to account for the saturation
of the charge carriers at increasing doping levels.
\cite{Guerrero_82} It is interesting to compare this model with
the latest experimental data. The mass action law for the reaction
(\ref{Cluster_Reaction_Tsai}) at the diffusion temperature yields
\cite{Tsai_80}

\begin{equation} \label{MAL_Tsai}
C^{Cl}=A^{Cl}nC^{3},
\end{equation}

\noindent where $C^{Cl}$ and $n$ are the concentrations of
clusters and electrons at the annealing temperature, respectively;
$A^{Cl}$ is the constant of local thermodynamic equilibrium. After
cooling, all clusters become neutral and $C\approx n_{R}$ if $C\gg
n_{i}$. Here $n_{R}$ is the concentration of electrons at room
temperature. It follows from the assumption of local charge
neutrality that the electron concentration at the annealing
temperature for $C\gg n_{i}$ is

\begin{equation} \label{Electron_equation_Tsai}
n\approx C+2C^{Cl}=C+2A^{Cl}nC^{3}.
\end{equation}

Solving this equation, Tsai et al. \cite{Tsai_80} obtained

\begin{equation} \label{Electron_Tsai}
n=\frac{C}{1-2A^{Cl}C^{3}}
\end{equation}

\noindent and

\begin{equation} \label{Total_Tsai}
C^{T}=C+\frac{3A^{Cl}C^{4}}{1-2A^{Cl}C^{3}}=\frac{C+A^{Cl}C^{4}}{1-2A^{Cl}C^{3}}.
\end{equation}

From Eq.(\ref{Total_Tsai}) it is inferred that under condition

\begin{equation} \label{Csat_Tsai}
C=C_{sat}=\frac{1}{\sqrt[3]{2A^{Cl}}}
\end{equation}

\noindent both the concentration of arsenic atoms incorporated
into clusters and the total concentration of arsenic atoms
approach infinity. The value $n_{Re}=C_{sat}$ is interpreted in
\cite{Tsai_80} as a maximum level of electron concentration. From
the experimental data it is seen that a maximum equilibrium
carrier concentration at the annealing temperature in the arsenic
implanted layers may be described by the following relation
\cite{Solmi_01}

\begin{equation} \label{Electron_Solmi}
n_{e}=1.3\times10^{11}\exp\left(-\frac{0.42 eV}{k_{B}T}\right).
\end{equation}

Using Eqs.\;(\ref{Electron_Solmi}) and (\ref{Csat_Tsai}), the
value of $A^{Cl}$ may be derived to be

\begin{equation} \label{Constant_Tsai}
A^{Cl}=\frac{1}{2n_{e}^{3}}.
\end{equation}

To illustrate, $n_{e}=3.57\times10^{8}$ $\mu $m$^{-3}$ at a
temperature of 1050\,$^{\circ}$C and $A^{Cl}=1.10\times10^{-26}$
$\mu $m$^{9}$.

The values of $n_{R}$ as a function of the total arsenic
concentration calculated by the model of  Tsai et al.
\cite{Tsai_80} are shown in Fig. 1. The experimental values of
$n_{R}$ based on the data from \cite{Solmi_98} is also given for
comparison. As seen from Fig. 1, saturation of the electron
concentration for the model \cite{Tsai_80} is observed at the
extremely high total As concentrations, whereas the experimental
data show that the carrier concentration reaches its maximum at
$C^{T}\approx 8\times10^{8}$ $\mu $m$^{-3}$. Thus, the model
\cite{Tsai_80} is inconsistent with the experimental data.
Moreover, as follows from the experimental data of Solmi and
Nobili, \cite{Solmi_98} the value of $n_{e}$ obtained at room
temperature corresponds to the peculiarity of the curve for
diffusivity as a function of the total arsenic concentration.
Because this feature characterizes impurity diffusion at the
annealing temperature, it was concluded in \cite{Solmi_98} that
$n_{e}$ had a physical meaning. Thus, the model \cite{Tsai_80}
that includes cluster neutralization during cooling is also
inconsistent with the experimental data for arsenic diffusivity.
Indeed, the electron concentration at the annealing temperature
calculated by this model and associated with the peculiarity of
the diffusivity is three times greater than $n_{e}$.

\begin{figure}[ht]
\centering {
\begin{minipage}[ht]{8.6 cm}
{\includegraphics[ scale=0.74]{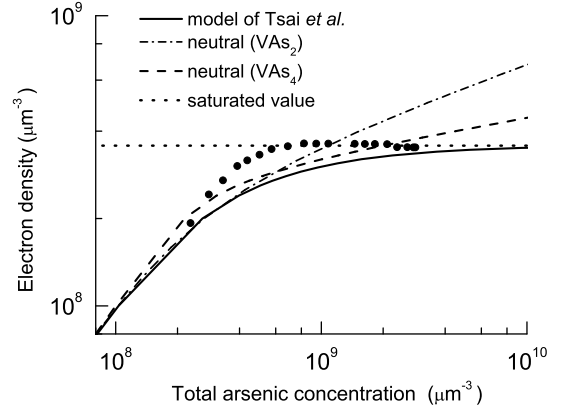}}

\end{minipage}
}\caption{Calculated electron concentration against the total
arsenic concentration for different models of clustering: solid
line --- model of Tsai et al. (Ref. \cite{Tsai_80}), dash-dotted
line --- neutral clusters $\text{VAs}_{2}$, and dashed line ---
neutral clusters $\text{VAs}_{4}$. The experimental data (circles)
are taken from Solmi and Nobili \cite{Solmi_98} for the diffusion
at a temperature of 1050\,$^{\circ}$C. \label{As_Cluster1}}
\end{figure}

Fig. 1 presents the functions $n=n(C^{T})$ calculated for the
cases when neutral clusters $VAs_{2}$ and $VAs_{4}$ are formed. In
these cases, saturation of the electron density is not observed and
the calculations do not agree with the experimental data. The
absence of saturation for the electron concentration corresponds
to the first-principles calculations in \cite{Berding_98} (Fig. 2)
and \cite{Berding_Sher_98} (Fig. 3). As seen from these figures,
no saturation is observed at arsenic concentrations when $VAs_{4}$
clustering is prevailing.

The reactions

\begin{equation} \label{Reaction_Guerrero}
m\text{As}^{+}+k e^{-}\rightleftarrows \text{As}_{m}^{r}
\end{equation}

\noindent were used in the basic model for the clustering of As in
silicon proposed by Guerrero et al. \cite{Guerrero_82} Here $m$
and $k$ are the numbers of arsenic atoms and electrons
participating in the clustering process; $r$ is the electric
charge of a cluster. The function $C=C(C^{T})$ was investigated by
means of the mass action law combined with the charge conservation
law for the cluster formation reaction on the assumption of local
charge neutrality

\begin{equation} \label{LCN_Guerrero}
n=C+rC^{Cl}=C+rA^{Cl}C^{m}n^{k}.
\end{equation}

It was supposed in \cite{Guerrero_82} that the charge of clusters
 becomes zero when the sample is cooled down to room
temperature. Therefore, $C$ is interpreted as the concentration of
electrically active arsenic at room temperature, which equals
the electron density $n$ in the sample at room temperature. Four
main conclusions have been drawn from the basic model:

1. When the electrons do not participate in the reaction
(\ref{Reaction_Guerrero}), i.e., $k=0$, or the clusters are
electrically neutral at high temperature, i.e., $r=0$, $C$ is a
monotonically increasing function of $C^{T}$ and also there is no
finite limit to $C$ and electron density $n$ at room temperature.

2. If the clusters are positively charged at a high temperature
($r\geq 1$) and exactly one electron takes part in the reaction
(\ref{LCN_Guerrero}), i.e., $k=1$, $C$ is a monotonically
increasing function of $C^{T}$, which approaches a finite
``saturation'' value $C_{max}$ with increasing $C^{T}$.

As can be see from the reaction (\ref{Cluster_Reaction_Tsai}), the
model of Tsai et al. \cite{Tsai_80} satisfies these conditions. On
the other hand, from our calculations for this model (see Fig. 1),
it follows that saturation of the electron concentration is
observed at extremely high total As concentrations. It means that
the saturation value obtained in such a manner is inconsistent
with the experimental data.

3. Provided that the clusters are positively charged at high
temperatures ($r\geq 1$) and there are two or more electrons
taking part in the reaction (\ref{Reaction_Guerrero}), i.e.,
$k\geq 2$, $C$ increases with increasing $C^{T}$ only up to a
finite value $C^{T}_{max}$. A maximum concentration value
$C=C_{max}$ is observed for $C^{T}=C^{T}_{max}$. With further
increase in the total arsenic concentration $C^{T}>C^{T}_{max}$
the concentration of the substitutionally dissolved arsenic atoms
$C$ and hence electron density $n$ at room temperature, are
decreased.

This behavior of $n$ is at variance with the experimental data,
because experimentally measured $n$ is approximately constant
 with increasing $C^{T}$ (see Fig. 1).

4. If the clusters are negatively charged at high temperatures
($r\leq -1$), then $C=C(C^{T})$  behaves as in case 1, i.e.,
$C$, and hence electron density $n$ at room temperature,
monotonically increase without limit for increasing  $C^{T}$. This
behavior of $n$ also conflicts with the experimental data.

In what follows, the basic model is extended to take into account
the different charge states of vacancies participating in the
cluster formation. \cite{Guerrero_82} It is shown that the main
conclusions from the basic model are valid for the case when
vacancy is involved in the formation of clusters. Thus, the
generalized model of arsenic clustering proposed in
\cite{Guerrero_82} is not in agreement with the experimental data
for all possible cases of the cluster formation considered in this
model. In our opinion, this contradiction is due to the assumption
that charged clusters become neutral during cooling. As follows
from, \cite{Solmi_98} the hypothesis that deactivation is taking
place during cooling of the substrates contradicts the
experimental data.

Evidently, the models of arsenic clustering should give an
adequate explanation not only for saturation of the electron
density, but also for the plateau formed on the carrier
concentration profile after annealing of the ion implanted layers.
\cite{Solmi_98,Solmi_01} As follows from the above analysis, there
is no model satisfying these requirements.

The performed analysis enables us to formulate the purpose of this
study: the development of a more adequate model %for arsenic clustering
to explain the plateau phenomenon of the charge carrier profile at
high concentration arsenic diffusion.

%For example, at higher temperatures concentration of isolated
%arsenic atom in the lattice $AsSi_{4}$ prevails on the
%concentration of $VAs_{4}$.

\section{\textbf{Model}}

To develop a model for arsenic clustering that can describe the
saturation phenomenon of the electron density, we assume that not
only neutral, but also negatively charged, arsenic clusters are
formed. The possibility for the formation of the negatively
charged arsenic clusters is confirmed by the calculations in.
\cite{Berding_98,Mueller_03} It has been shown by Berding et al.
\cite{Berding_98} that clusters $VAs_{3}$ and $VAs_{2}$ have one
and two acceptor levels, respectively. The calculations carried
out by Mueller et al. \cite{Mueller_03} also show that for the
electron densities $n>10^{4}$ $\mu \text{m}^{-3}$ the majority of
complexes $\text{VAs}_{2}$ will be negatively charged. According
to, \cite{Mueller_03} the complex $\text{VAs}_{3}$ exhibits an
acceptor level. This cluster is singly ionized at elevated Fermi
levels, resulting in a loss of four mobile carriers for the
formation of a $\text{VAs}_{3}$ aggregate.

It is interesting that saturation of the electron density is
observed in the layers of GaAs heavily doped by silicon. To
explain this phenomenon, it was supposed (see, for example
\cite{Greiner_84,Velichko_93}) that silicon atoms in heavily doped
GaAs are dissolved substitutionally not only in the Ga sublattice
(donor $\text{Si}_{\text{Ga}}^{+}$), but also in the As sublattice
(acceptor $\text{Si}_{\text{As}}^{-}$).  It was supposed in
\cite{Adler_95} that a maximum electrical conductivity in Sb-doped
silicon is limited by the defect formation at high dopant
concentrations. The Sb-vacancy pairs were considered as defect
dominating at high doping levels. Also, it was supposed that this
defect acts as an electron acceptor and is responsible for
saturation of charge carriers at high doping levels. Although a
mechanism of such a saturation of the electron density in silicon
heavily doped by Sb is different from that in As-doped silicon,
\cite{Solmi_01} the idea of the limited electron density due to
compensation by the acceptors is a very fruitful one. For example,
it has been supposed in \cite{Velichko_Dobrushkin_Pakula_05} that
phosphorus clusters are negatively charged. Due to this
assumption, saturation of the electron density during high
concentration phosphorus diffusion was explained.

As distinct from the models of arsenic clustering considered in,
\cite{Tsai_80,Guerrero_82} we assume that arsenic clusters have
the same negative charge both at high diffusion temperatures and
at room temperature. As follows from the analysis of the
experimental data and theoretical investigations, the structure of
arsenic clusters under diffusion temperature is still not
conclusively established. Therefore, it is assumed that a point
defect $\text{D}_{1}$ can be involved in the cluster formation and
another defect $\text{D}_{2}$ can be generated during clustering.
In this case, a reaction for the formation and dissolution of
clusters can be written as

\begin{equation}\label{Reaction_Basic}
m\text{As}^{+}  + m_{1}\text{D}^{r_{1}}+ ke^{ -}
{\longleftrightarrow} \Phi^{r_{Cl} } + m_{2}\text{D}^{r_{2}} ,
\end{equation}

\noindent where A is the substitutionally dissolved arsenic atom
$\text{As}^{+}$; $e^{-}$ is the electron; $\Phi$ is the cluster
formed; $m_{1}$ and $m_{2}$ are respectively the numbers of
defects $\text{D}_{1}$ and $\text{D}_{2}$ participating in the
cluster formation; $r_{1}$, $r_{2}$, and $r_{Cl}$  are the charge
states of defect $\text{D}_{1}$, defect $\text{D}_{2}$, and
cluster $\Phi^{r_{Cl} }$, respectively.

The charge conservation law for the chemical reaction
(\ref{Reaction_Basic}) is of the form

\begin{equation} \label{CCL_Basic}
m + m_{1} z_{1} - k = z_{Cl}+ m_{2} z_{2}  ,
\end{equation}

\noindent where $z_{1}$ and  $z_{2}$ are the charges of defects
$\text{D}_{1}$ and $\text{D}_{2}$, respectively; $z_{Cl}$ is the
charge of cluster $\Phi^{r_{Cl} }$ in terms of the elementary
charge.

The mass action law for reaction (\ref{Reaction_Basic}) yields

\begin{equation} \label{MAL_Basic}
K_{L} \;C^{\DS{m}} \;(C_{D1}^{r_{1}} )^{\DS{m_{1}}}
\;\chi^{\DS{k}}n_{i}^{\DS{k}}  = K_{R} \;C_{\Phi}
\;(C_{D2}^{r_{2}} )^{\DS{m_{2}}}  ,
\end{equation}

\noindent where $K_{L}$ is the rate of chemical reaction
(\ref{Reaction_Basic}) in the forward direction; $K_{R}$ is the
rate of this reaction in the opposite direction; $C$ is the
concentration of impurity atoms participating in reaction
(\ref{Reaction_Basic}) ; $C_{D1}^{r_{1}}$  is the concentration of
defects in the charge state $r_{1}$ participating in the cluster
formation; $C_{D2}^{r_{2}}$  is the concentration of defects in
the charge state $r_{2}$ generated during clustering; $C_{\Phi}$
is the concentration of the clusters in the charge state $r_{Cl}$;
$\chi$ is the electron density normalized to the concentration of
the intrinsic charge carriers in a semiconductor during diffusion
$n_{i}$.

Due to a high dopant concentration, the approximation of local
charge neutrality can be used for the evaluation of $\chi$

\begin{eqnarray} \label{chi}
\chi = \DS{\frac{n}{n_{i}}} &=& \frac{1}{2n_{i}} \biggl[ C+ z_{Cl}
\; C_{\Phi} - N_{B} \nonumber \\
    &+& \left. \sqrt {\left( C +z_{Cl} \; C_{\Phi} - N_{B}
\right)^{2} + 4n_{i}^{2}}\, \right]\, ,
\end{eqnarray}

\noindent where $N_{B}$ is the summarized concentration of
acceptors; $n_{ie}$ is the effective concentration of the
intrinsic charge carriers at a diffusion temperature calculated
for a high doping level.

It follows from equation (\ref{Reaction_Basic})  that the
concentration of impurity atoms incorporated into clusters $C^{A}
= mC_{\Phi}$. Moreover, considering highly mobile electrons, the
mass action law for the reaction of defect conversion between the
charge states is valid

\begin{equation} \label{MAL_Electron_1}
C_{D1}^{r_{1}}  = h_{D1}^{r_{1}} \;n^{\DS{-z_{1}}} \,
C_{D1}^{\times} = h_{D1}^{r_{1}} \, n_{i}^{\DS{ - z_{1}}} \;\chi
^{\DS{ - z_{1}}} \, C_{D1}^{\times} \quad {  ,}
\end{equation}

\begin{equation} \label{MAL_Electron_2}
C_{D1}^{r_{2}}  = h_{D2}^{r_{2}} \;n^{\DS{-z_{2}}} \,
C_{D2}^{\times} = h_{D2}^{r_{2}} \, n_{i}^{\DS{ - z_{2}}} \;\chi
^{\DS{ - z_{2}}} \, C_{D2}^{\times} \quad {  ,}
\end{equation}

\noindent where $h_{D1}^{r_{1}}$ and $h_{D2}^{r_{2}}$ are the
constants for the local thermodynamic equilibrium in the reactions
when neutral defects $\text{D}_{1}^{\times}$ and
$\text{D}_{2}^{\times}$ are converted into charge states $r_{1}$
and $r_{2}$, respectively.

Substituting (\ref{MAL_Electron_1}) and (\ref{MAL_Electron_2})
into (\ref{MAL_Basic}) and taking into consideration that $C_{AC}
= mC_{Cl}$, we obtain

\begin{equation} \label{CA}
C_{AC} = K\; \tilde {C}_{D} \;\chi ^{\displaystyle{(
\;k\;+r_{2}m_{2}-r_{1}m_{1})}} \;C^{\displaystyle{m}} ,
\end{equation}

\noindent where

\begin{equation} \label{Constant}
K = \frac{m\;K_{L} \,(h_{D1 \DS{r_{1}}})^{\DS{m_{1}}} \,
n_{i}^{\DS{(k+r_{2}m_{2}-r_{1}m_{1})}} \,(C_{D1
\DS{\times}}^{eq})^{\DS{m_{1}}}} {K_{R} \,(h_{D2
\DS{r_{2}}})^{\DS{m_{2}}} \,(C_{D2 \DS{\times}}^{eq})^{\DS{m_{2}}}
\, },
\end{equation}

\begin{equation} \label{Rel_Defect}
\tilde {C}_{D}=\frac{(\tilde {C}_{D1})^{\DS{m_{1}}}}{(\tilde
{C}_{D2})^{\DS{m_{2}}}} \, , \quad \tilde {C}_{D1}= \frac{{C}_{D1
\DS{\times}}}{{C}_{D1 \DS{\times}}^{eq}} \, ,\quad \tilde
{C}_{D2}= \frac{{C}_{D2 \DS{\times}}}{{C}_{D2 \DS{\times}}^{eq}}
\, .
\end{equation}

\noindent Here $C_{AC}$ is the concentration of clustered arsenic
atoms and $C_{T} = C + C_{AC}$ is the total arsenic concentration.
The parameter $K$ has a constant value depending on the
temperature of diffusion. This value can be extracted from the
best fit to the experimental data. The quantities ${C}_{D1
\DS{\times}}^{eq}$ and ${C}_{D2 \DS{\times}}^{eq}$ are the
equilibrium concentrations of corresponding defects in the neutral
charge state.

\section{\textbf{Results of calculations}}

Analysis of Eqs. (\ref{chi}) and (\ref{CA}) shows that saturation
of the electron density will be observed for the formation of
singly negatively charged clusters incorporating one arsenic atom
and a lattice defect. The calculations of the functions $C =
C(C^{T})$, $C^{A} = C^{A}(C^{T})$, and $n = n(C^{T} )$ for this
case are presented in Fig. 2. For comparison with the experimental
data, \cite{Solmi_98} the processing temperature was chosen to be
1050\,$^{\circ}$C ($n_{i} $ = 1.1665$\times10^{7}$ $\mu$m$^{-3}$).
To calculate the functions $C = C\left( C^{T} \right)$,
$C^{A}(C^{T})$, and $n(C^{T})$, a system of nonlinear equations
(\ref{chi}), (\ref{CA}) was solved numerically by Newton's method.
In the case of singly negatively charged clusters
$(\text{DAs})^{-}$ incorporating a lattice defect and one arsenic
atom we adopt $r_{Cl}$ = -1 and $m_{1} $ = 1. Eq.
(\ref{CCL_Basic}) for $m$ = 1 yields that $- z_{1}m_{1} -
z_{2}m_{2} + k$ = 2, and the system (\ref{CA}), (\ref{chi}) is of
the form:

\begin{equation} \label{CA1}
C^{A} = K\;\tilde {C}_{D} \;\chi ^{2}\;C \; ,
\end{equation}

\begin{eqnarray} \label{chi1}
\chi &=& \frac{1}{2n_{i}} \biggl[ C - K \tilde{C}_{D} \chi^{2}C -
N_{B}+ \nonumber \\
     &+& \left.  \sqrt {\left( C - K \tilde {C}_{D} \chi ^{2}C - N_{B}
\right)^{2} + 4n_{ie}^{2}}\quad\right] {  .}
\end{eqnarray}

As can be seen from Fig. 1, the concentration of charge carriers
in the saturation region is approximately equal to the value
$n_{e}=$ 3.5665$\times $10$^{8}$ $\mu$m$^{-3}$. In Fig. 2 for
$C^{T}\rightarrow+\infty$, the electron concentration reaches
saturation at the value $n \sim $ 3.765$\times $10$^{8}$
$\mu$m$^{-3}$ associated with the plateau of the charge carrier
profile given in Fig. 1. The value of parameter $K\tilde{C_{D}}$
for this case was chosen as $ 1.0698\times
 10^{{ -4}}$ a. u. to provide a fit to the experimental value of
$n_{e}$. As seen from Fig. 2, the function $n=n(C^{T})$ is similar
to that predicted by the model of Tsai et al. \cite{Tsai_80}
Saturation of the electron density occurs at extremely high values
of the total impurity concentration $C^{T}$, which is in conflict
with the experimental data.

%\begin{figure}
%\includegraphics{}
%\caption{\label{}}
%\end{figure}

\begin{figure}[ht]
\begin{minipage}[ht]{8.6 cm}
{\includegraphics[ scale=0.74]{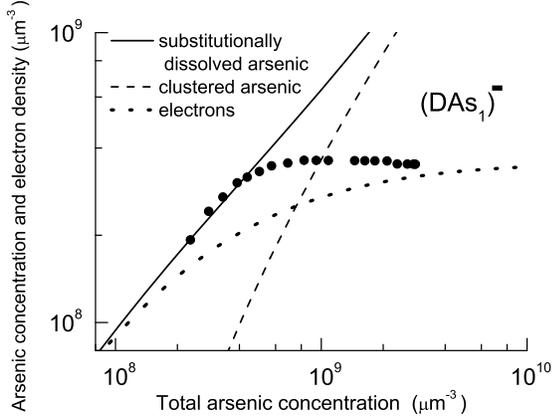}}
\end{minipage}

\caption{Calculated concentrations of substitutionally dissolved
arsenic atoms (solid line), clustered arsenic atoms (dashed line),
and electron density (dotted line) against the total dopant
concentration for the formation of singly negatively charged
clusters incorporating one arsenic atom. The measured electron
density (circles) is taken from Solmi and Nobili \cite{Solmi_98}
for the diffusion at a temperature of 1050\,$^{\circ}$C.
\label{As_Cluster2}}
\end{figure}

In Fig. 3a the system of nonlinear equations (\ref{chi}),
(\ref{CA}) is solved numerically for the formation of singly
negatively charged  clusters $(DAs_{2})^{-}$ incorporating a
lattice defect and two arsenic atoms ($r_{Cl} $ = -1, $m_{1} $ =
1, and $m$ = 2). Then, from (\ref{CCL_Basic}), (\ref{chi}), and
(\ref{CA}) it follows that:

\begin{equation} \label{CA2}
C^{A} = K\;\tilde {C}_{D} \;\chi ^{3}\;C^{2} \, ,
\end{equation}

\begin{eqnarray} \label{chi2}
\chi &=& \frac{1}{2n_{i}} \Biggl[ C - \DS{\frac{1}{2}}K
\tilde{C}_{D} \chi^{3}C^{2} - N_{B}+ \nonumber \\
     &+& \left.  \sqrt {\left( C - \DS{\frac{1}{2}}K \tilde {C}_{D} \chi ^{3}C^{2} - N_{B}
\right)^{2} + 4n_{ie}^{2}}\, \right] .
\end{eqnarray}

The calculations for $n=n(C^{T})$ show that, with the increasing
total concentration of impurity atoms $C^{T}$, the concentration of
charge carriers $n$  reaches its maximum value $n_{max}$
=3.750$\times 10^{8}$ $\mu$m$^{-3}$ at  $C^{T}_{\max} \sim
1.634\times$10$^{9}$ $\mu$m$^{-3}$ and then decreases
monotonically. At maximum the concentration of the
substitutionally dissolved arsenic atoms $C_{\max}$ is
$7.943\times$10$^{8}$ $\mu$m$^{-3}$. The parameter
$K\tilde{C_{D}}$ used in this calculation is equal to
4.0$\times$10$^{-14}$ $\mu$m$^{3}$. The calculated curve agrees
with the experimental function $n=n(C^{T})$ much better than those
obtained by the model of Tsai et al. \cite{Tsai_80} and the models
of neutral $(\text{DAs}_{2})$ and neutral $(\text{DAs}_{4})$
clusters.

\begin{figure}[ht]
\begin{minipage}[ht]{8.6cm}
{\includegraphics[ scale=0.74]{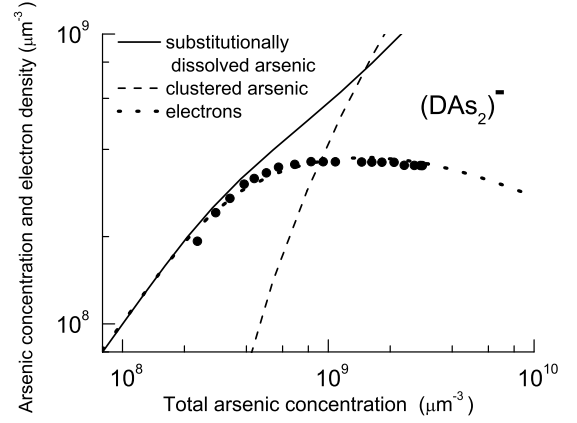}} \centerline{(a)}
\end{minipage}

\begin{minipage}[t]{8.6cm}
{\includegraphics[ scale=0.74]{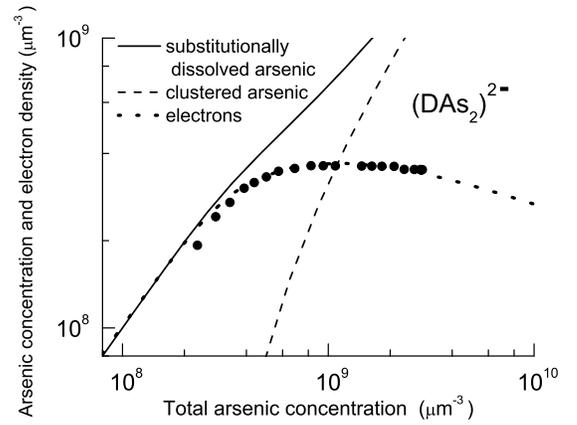}}

\centerline{(b)}
\end{minipage}

\caption{Calculated concentrations of substitutionally dissolved
arsenic atoms (solid line), clustered arsenic atoms (dashed line),
and electron density (dotted line) against the total dopant
concentration for the formation of  singly (a) and doubly (b)
negatively charged clusters incorporating two arsenic atoms. The
measured electron density (circles) is taken from Solmi and Nobili
\cite{Solmi_98} for the diffusion at a temperature of
1050\,$^{\circ}$C. \label{As_Cluster3a-b}}
\end{figure}

The calculations of $n=n(C^{T})$ and  $C^{A}=C^{A}(C^{T})$ were
also performed for the clusters $(\text{DAs}_{2})^{2-}$,
$(\text{DAs}_{3})^{-}$, $(\text{DAs}_{3})^{2-}$,
$(\text{DAs}_{4})^{-}$, and $(\text{DAs}_{4})^{2-}$. And similar
behavior of  $n=n(C^{T})$ was observed for all these cases. Also,
the concentration of charge carriers increases with increasing
$C^{T}$, reaches a maximum, and then decreases monotonically. This
decrease is more pronounced with an increased number of arsenic atoms
in the cluster. Besides, the position of maximum carrier
concentration is shifted to smaller dopant concentrations with
increases in the negative charge of the cluster.

It follows from comparison of the calculated distributions
$n(C^{T})$ with the experimental data that the best fit is
observed for the doubly negatively charged cluster
$(\text{DAs}_{2})^{2-}$ (see Fig. 3b). To obtain the curves shown
in Fig. 4, we solve the system (\ref{chi}), (\ref{CA}) having the
following form for the clusters $(\text{DAs}_{2})^{2-}$ \, :

\begin{equation} \label{CA22}
C^{A} = K\;\tilde {C}_{D} \;\chi ^{4}\;C^{2} \; ,
\end{equation}

\begin{eqnarray} \label{chi22}
\chi &=& \frac{1}{2n_{i}} \Biggl[ C - K \tilde{C}_{D}
\chi^{4}C^{2} - N_{B}+ \nonumber \\
     &+& \left.  \sqrt {\left( C - K \tilde {C}_{D}
      \chi ^{4}C^{2} - N_{B}
\right)^{2} + 4n_{ie}^{2}} \, \right] \, .
\end{eqnarray}

At maximum value of $n_{max}=3.694\times$10$^{8}$ $\mu$m$^{-3}$,
the concentration of the substitutionally dissolved arsenic atoms
$C_{\max}$ is equal to $7.943\times$10$^{8}$ $\mu$m$^{-3}$. The
calculated curve $n=n(C^{T})$ agrees well with the experimental
data, although only one fitting parameter $K\tilde{C_{D}}=$
0.67$\times$10$^{-15}$ $\mu$m$^{3}$ has been used. Full agreement
is reached in the region of the transition to saturation of the
electron density $n$. This means that two arsenic atoms are
incorporated in the cluster, and this arsenic cluster is doubly
negatively charged, at least at the total dopant concentrations
$C^{T}\leq$ 1.0$\times$10$^{9}$ $\mu$m$^{-3}$. Minor differences
in values of the theoretical curve and experimental data for
$C^{T}>$ 2.3$\times$10$^{9}$ $\mu$m$^{-3}$ can arise due to heavy
doping effects (change of the silicon zone structure, changes in
the constants of forward $K_{L}$ and backward $K_{R}$ reactions,
etc). It is clear from expressions (\ref{chi}), (\ref{CA}) that
variation of the defect concentration $\tilde {C}_{D}$ in the
region of high impurity concentrations also influences the
concentration of impurity atoms incorporated in the clusters and,
hence, the concentration of charge carriers. A change of $\tilde
{C}_{D}$ due to generation (absorption) of the defects in the
region with high arsenic concentration is quite possible. For
example, such a generation can occur due to the lattice expansion.
Besides, it is possible that a high doping level leads to error in
the determination of the experimental values for electron density,
and the data for $C^{T}>$ 2.3$\times$10$^{9}$ $\mu$m$^{-3}$ may be
incorrect.

Figs. 4 and b show the functions $C = C(C^{T})$, $C^{A} =
C^{A}(C^{T})$, and $n = n(C^{T})$ calculated for clusters
$(\text{DAs}_{3})^{-}$ and $(\text{DAs}_{4})^{-}$, respectively.
As can be seen from these figures, fitting to the experimental
data is worse in comparison with Fig. 3b, especially for the total
concentrations greater than 2.0$\times$10$^{9}$ $\mu$m$^{-3}$.

\begin{figure}[ht]
\begin{minipage}[ht]{8.6cm}
{\includegraphics[ scale=0.74]{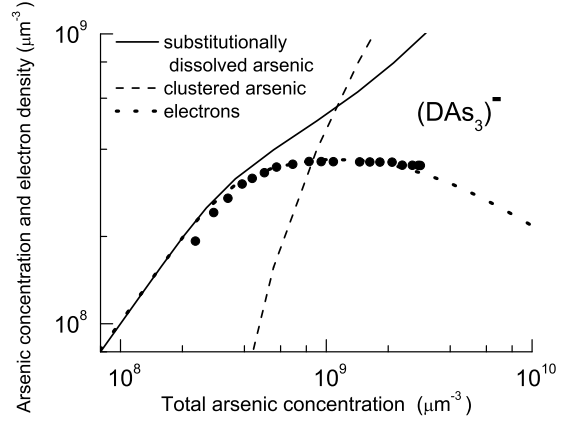}} \centerline{(a)}
\end{minipage}

\begin{minipage}[ht]{8.6 cm}
{\includegraphics[ scale=0.74]{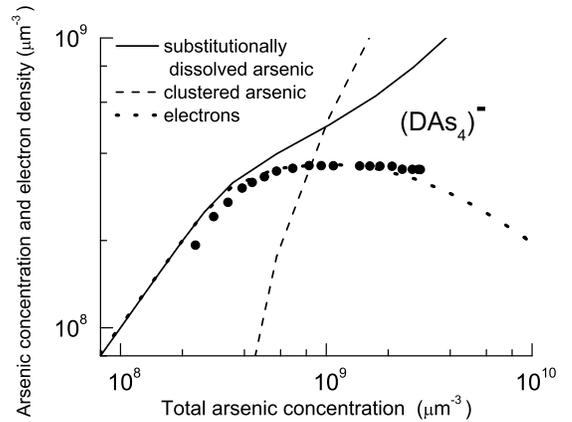}} \centerline{(b)}
\end{minipage}

\caption{Calculated concentrations of substitutionally dissolved
arsenic atoms (solid line), clustered arsenic atoms (dashed line),
and electron density (dotted line) against the total dopant
concentration for the formation of singly negatively charged
clusters incorporating three (a) and four (b) arsenic atoms. The
measured electron density (circles) are taken from Solmi and
Nobili \cite{Solmi_98} for the diffusion at a temperature of
1050\,$^{\circ}$C. \label{As_Cluster4a-b}}
\end{figure}

Note that multiple clustering can occur, too, and different clusters
can coexist in the equilibrium state. For example, the clusters
incorporating different numbers of arsenic atoms
$(\text{DAs}_{1})^{-}$ and $(\text{DAs}_{2})^{-}$ or clusters in
different charge states $(\text{DAs}_{2})^{\times}$,
$(\text{DAs}_{2})^{-}$, and $(\text{DAs}_{2})^{2-}$ can coexist.
As follows from the calculated curves for the electron density,
when clusters $(\text{VAs}_{1})^{-}$ (Fig. 2) and
$(\text{VAs}_{2})^{\times}$ (Fig. 1) are formed, it may be
expected that the fall of the electron density in the plateau region
will be less pronounced in the case of multiple clustering.
However, this problem requires further investigation.

\section{\textbf{Conclusions}}

Based on analysis of the experimental results and theoretical
calculations for clustering of arsenic atoms in silicon, it was
shown that the developed clustering models fail to explain the
available experimental data, especially the effect of electron
density saturation at high arsenic concentrations. Therefore, a
new model of arsenic clustering was proposed and analyzed. The
main feature of the proposed model is the assumption that
negatively charged arsenic complexes play a dominant role in the
clustering process. To confirm this assumption, the concentration
of the impurity atoms incorporated in clusters and electron
density were calculated as a function of the total arsenic
concentration at a temperature of 1050\,$^{\circ}$C. Different
cases of the formation of negatively charged clusters
incorporating a point defect and one arsenic atom
$(\text{DAs}_{1})^{-}$, $(\text{DAs}_{1})^{2-}$ or more arsenic
atoms $(\text{DAs}_{2})^{-}$, $(\text{DAs}_{2})^{2-}$,
$(\text{DAs}_{3})^{-}$, $(\text{DAs}_{3})^{2-}$,
$(\text{DAs}_{4})^{-}$, and $(\text{DAs}_{4})^{2-}$ were
investigated. It was shown that in the case of
$(\text{DAs}_{1})^{-}$ the concentration of charge carriers
reaches saturation with increase in the total arsenic
concentration and concentration of the substitutionally dissolved
impurity atoms. However, saturation was observed for very high
values of the total arsenic concentration, conflicting with the
experimental data. In the cases of doubly negatively charged
cluster $(\text{DAs}_{1})^{2-}$ and clusters incorporating more
than one arsenic atom, the electron density reached its maximum
value, then  decreased monotonically and slowly with an increase
in the total arsenic concentration. The electron density
calculated for the formation of $(\text{DAs}_{2})^{2-}$ clusters
agrees well with the experimental data and confirms the conclusion
that two arsenic atoms participate in the cluster formation. Minor
difference between the theoretical curve and experimental values
for $C^{T}>$ 2.3$\times$10$^{9}$ $\mu$m$^{-3}$ may be explained by
various heavy doping effects or uncertainty due to the
experimental errors.

Thus, the plateau on the experimental distribution of the charge
carriers, characterized by a practically constant value of the
electron density in the region of high doping level, may be
attributed to the negatively charged clusters
$(\text{DAs}_{2})^{2-}$. The parameters describing arsenic
clustering at a temperature of 1050\,$^{\circ}$C were determined
by fitting the calculated values of the electron density to the
experimental data. The proposed model of clustering by the
formation of the negatively charged $(\text{DAs}_{2})^{2-}$ system
gives the best fit to the experimental data among all the
present-day models and may be used in simulation of high
concentration arsenic diffusion.

\end{document}